# Electronic Raman Scattering in Suspended Semiconducting Carbon Nanotubes

Yuecong Hu (胡悦聪),[1,*] Shaochuang Chen (陈少闯),[1,*] Xin Cong (从鑫),[2] Sida Sun (孙斯达),[1] Jiang-bin Wu (吴江滨),[2] Daqi Zhang (张达奇),[1] Feng Yang (杨烽),[1] Juan Yang (杨娟),[1,†] Ping-heng Tan (谭平恒),[2] Yan Li (李彦)[1,‡]

[1] *Beijing National Laboratory for Molecular Sciences, Key Laboratory for the Physics and Chemistry of Nanodevices, State Key Laboratory of Rare Earth Materials Chemistry and Applications, College of Chemistry and Molecular Engineering, Peking University, Beijing 100871, China*

[2] *State Key Laboratory of Superlattices and Microstructures, Institute of Semiconductors, Chinese Academy of Sciences, Beijing 100083, China*

*These authors contributed equally to this work.

†To whom all correspondence should be addressed. yang_juan@pku.edu.cn

‡To whom all correspondence should be addressed. yanli@pku.edu.cn

# ABSTRACT

The electronic Raman scattering (ERS) features of single-walled carbon nanotubes (SWNTs) can reveal a wealth of information about their electronic structures, but have previously been thought to appear exclusively in metallic (M-) but not in semiconducting (S-) SWNTs. We report the experimental observation of the ERS features with an accuracy of ±1 meV in suspended S-SWNTs, the processes of which are accomplished via the available high-energy electron-hole pairs. The ERS features can facilitate further systematic studies on the properties of SWNT, both metallic and semiconducting, with defined chirality.

The electronic and phonon structures of the low-dimensional materials [1] are of vital importance in understanding their various properties. As a versatile tool, Raman spectroscopy has provided insight into those structures and the related properties [2,3]. Single-walled carbon nanotubes (SWNTs) are unique one-dimensional systems with abundant phonon [4,5] and electronic structures [6,7], which can be probed by the phonon Raman scattering and the electronic Raman scattering (ERS), respectively. The phonon Raman processes involve the inelastic light scattering by various phonon modes, and have been widely utilized in SWNT characterization [8-15]. The ERS processes, on the other hand, originate from the inelastic light scattering by a continuum of electron-hole (*e-h*) pairs. Consequently, the ERS processes can be resonantly enhanced at the excitonic transition energies ($E_{ii}$) of the SWNTs [16-19]. Due to the narrower bandwidth of the ERS features (30-50 meV) compared to the electronic transitions in regular electronic spectroscopies such as Rayleigh scattering [20,21] and optical absorption spectroscopies [22] (100-150 meV), the ERS spectra can provide direct and accurate information about the $E_{ii}$ of SWNTs, which are of great importance to nanotube-related science and technology.

Previously, the ERS processes have been exclusively reported in metallic SWNTs (M-SWNTs) and attributed only to the low-energy *e-h* pairs created across their linear electronic sub-bands near the Fermi level [16-19]. Therefore, the ERS features have been thought to appear exclusively in M-SWNTs but not in semiconducting SWNTs (S-SWNTs), which are more desired in many application fields such as nanoelectronics [23-27] and bioimaging [28-31]. This has restricted the understanding to the fundamental processes of ERS and limited their potential applications.

In this letter, we extend the ERS processes to all types of SWNTs and show that the ERS features are not only applicable to M-SWNTs but also can be observed in S-SWNTs. Our results support that the ERS processes are accomplished by a continuum of *e-h* pairs across the electronic sub-bands, both low-energy and high-energy. We report that the $E_{ii}$ ($M_{ii}$ for M-SWNTs and $S_{ii}$ for S-SWNTs) with an accuracy of ±1 meV can be directly obtained via the ERS spectra, compared to a typical accuracy of ±10 meV in conventional electronic spectroscopies [20-22]. The ERS features, which can be observed up to ~9000 cm$^{-1}$ (~1.1 eV) away from the excitation laser energy, reveal a wealth of information about the electronic structures of the SWNTs that can facilitate further systematic studies.

The electronic band structure of S-SWNTs differs from that of the M-SWNTs in the absence of a pair of linear electronic sub-bands arising from the gapless dispersion near the *K* point. Therefore, S-SWNTs do not possess the low-energy *e-h* pairs ($E_{e-h} < S_{11}$) across the linear electronic sub-bands but have high-energy *e-h* pairs with energy above $S_{11}$ ($E_{e-h} \geq S_{11}$). The ERS processes in S-SWNTs can be accomplished by a continuum of those high-energy *e-h* pairs. The electronic band structures of the first-order intravalley interactions of the ERS processes in S-SWNTs are schematically shown in Figure 1(a), and the corresponding density of states diagram is given in Figure 1(b). If the excitation laser energy ($E_L$) is set to be slightly higher than an $E_{ii}$, the ERS spectra can only be observed for M-SWNTs but not for S-SWNTs because only the low-energy *e-h* pairs matching the corresponding energy difference of $E_L - E_{ii}$ in M-SWNTs can contribute to this ERS process. No such low-energy *e-h* pairs are present in S-SWNTs to accomplish this ERS process. However, if $E_L$ is set to be sufficiently high so that the energy difference of $E_L - E_{ii}$ can match those available high-energy *e-h* pairs in S-SWNTs, we do expect to observe the ERS spectra in S-SWNTs that can be accomplished by the continuum of those high-energy *e-h* pairs. For instance, if $E_L \geq S_{11} + S_{22}$ is satisfied, an ERS feature can be anticipated at $S_{22}$ involving in the light scattering by the continuum of $e$-$h$ pairs near $E_{e\text{-}h} = E_L - S_{22} \geq S_{11}$. Similar ERS features are expected at higher $S_{ii}$ with an $E_L$ satisfying $E_L \geq S_{11} + S_{ii}$.

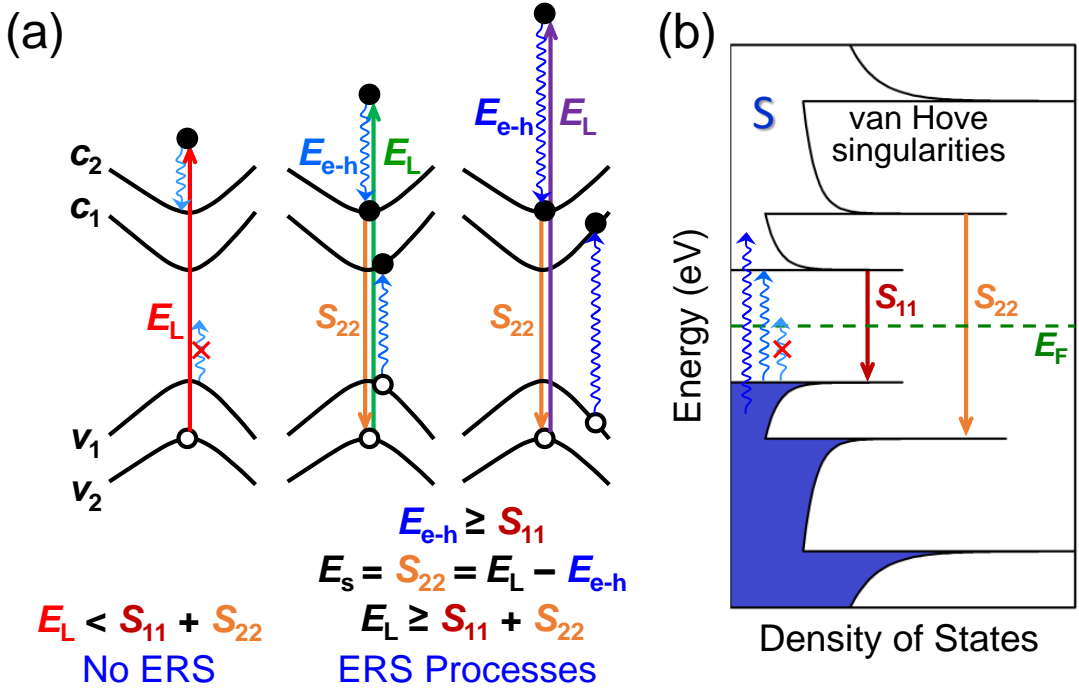

FIG. 1 Schematics of the ERS processes in S-SWNTs. (a) Electronic band structures of the first-order intravalley interactions of the ERS processes at different excitation lasers in an S-SWNT. Only high-energy $e$-$h$ pairs with $E_{e\text{-}h} \geq S_{11}$ are available in S-SWNTs. If $E_L < S_{11} + S_{22}$, no ERS is expected due to the absence of the low-energy $e$-$h$ pairs matching the corresponding energy difference of $E_L - E_{22}$. When $E_L \geq S_{11} + S_{22}$ is satisfied, the ERS processes are expected at $S_{22}$ by the continuum of $e$-$h$ pairs near $E_{e\text{-}h} = E_L - S_{22} \geq S_{11}$. $E_L$, $E_s$, and $E_{e\text{-}h}$ represent the energies of laser, scattered photon, and $e$-$h$ pair, respectively. (b) Density of states diagram of this S-SWNT with the corresponding $e$-$h$ pairs and electronic transitions.

The individual suspended SWNTs grown over the open slits (typical width of 65 μm) of SiO$_2$/Si substrate with Co catalysts via chemical vapor deposition [18,32] were used in this study. Figure 2(a) shows the Raman spectrum of a suspended S-SWNT at 532 nm excitation in a wide range of 100-7500 cm$^{-1}$. According to the radial breathing mode (RBM) at 187 cm$^{-1}$ observed with both 532 and 785 nm lasers and the photoluminescence (PL) peak at 0.799 eV with the 785 nm laser [Fig. 2(b)], we can determine its chirality as (15,2) using the (n,m) assignment program we have previously proposed (www.chem.pku.edu.cn/cnt_assign) [33]. From the (*n*,*m*) calculator we launched on the same website, we calculate that $S_{11}$ = 0.803 eV and $S_{22}$ = 1.522 eV for (15,2). The summation of $S_{11}$ + $S_{22}$ = 2.325 eV is slightly lower than the 532 nm laser energy ($E_L$ = 2.330 eV). Therefore, we expect an ERS feature at $E_{e\text{-}h} = E_L - S_{22}$ = 0.808 eV, which corresponds to ~6520 cm$^{-1}$. Experimentally, we observe two features in the 6000-7500 cm$^{-1}$ region: a phonon peak that can be assigned to the G+4D combinational mode at 6802 cm$^{-1}$ with a full width at half maximum (FWHM) of 99 cm$^{-1}$ (12 meV), and an ERS band at 6587 cm$^{-1}$ with a FWHM of 291 cm$^{-1}$ (36 meV). The experimental spectra agree with our expectation. Note that the combinational mode G+4D show significantly enhanced Raman intensity compared to the combinational modes of G+2D and 2G+2D. This can be explained by the scattered resonance condition [7,11] of its scattered photon energy with the $S_{22}$. Considering a spectral resolution of 0.7 meV (~6 cm$^{-1}$ with a 300 grooves/mm grating), an error introduced by random noise of 0.2 meV, and a spectral fitting error with different initial parameters of 1.3 meV, we estimate that the error in determining $S_{22}$ is ±1 meV. Therefore, we calculate an accurate value of $S_{22}$ = 1.513 eV from the ERS feature for this particular nanotube. Together with what we have reported previously for M-SWNTs [19], we conclude that $E_{ii}$ with a high accuracy of ±1 meV can be directly obtained via the ERS features for all types of SWNTs, both metallic and semiconducting.

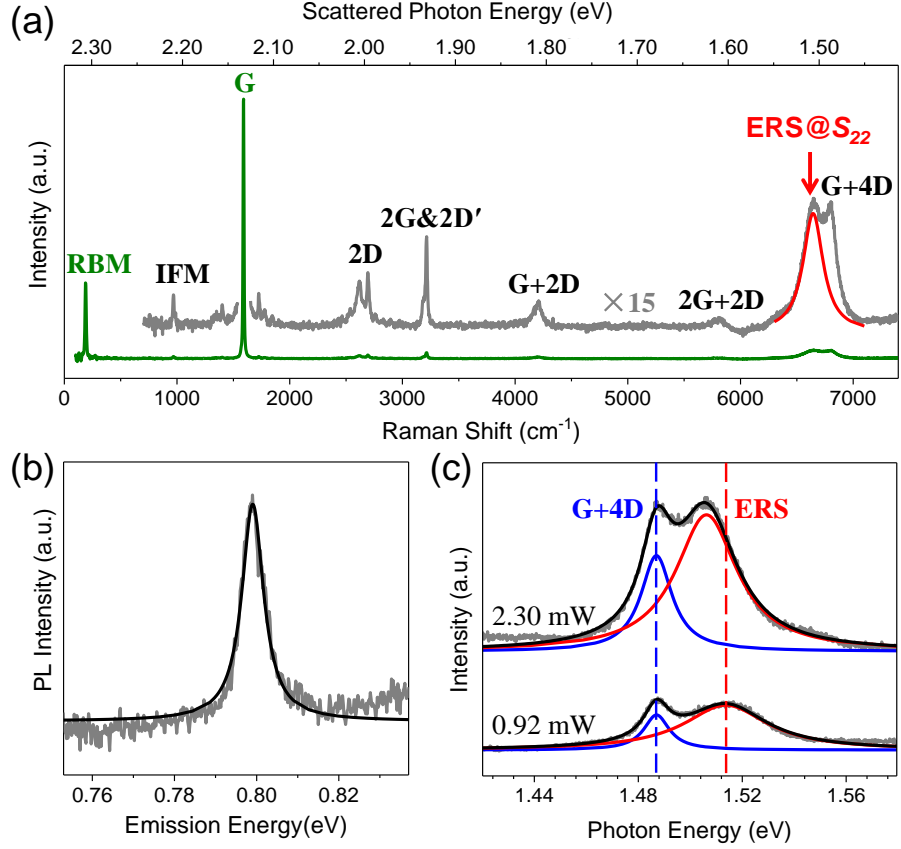

FIG. 2 Raman and PL spectra of a suspended (15,2) tube. (a) Wide range Raman spectra at 532 nm excitation, showing various phonon peaks and an ERS band corresponding to $S_{22}$. (b) PL spectra at 785 nm excitation. (c) Raman spectra at different powers of the 532 nm laser. The ERS band, the phonon peaks, and the PL peak are fitted with a Lorentzian line shape.

We intentionally heat this suspended (15,2) nanotube with different powers of the 532 nm laser [Fig. 2(c)] to reveal how temperature changes have affected both the ERS feature and the phonon peak. Compared to the spectrum collected at a low laser power (0.92 mW) showing negligible laser heating effect, at an increased laser power (2.30 mW) the ERS feature redshifts due to the thermal expansion of the nanotube and the consequently decreased transition energies whereas the G+4D phonon peak remains nearly unchanged. Further increase in the laser power leads to burning down of this nanotube. The laser heating effect on the ERS features of S-SWNTs are in good agreements with our previous data on the ERS features of M-SWNTs [18]. We find that suspended S-SWNTs are more vulnerable to increased laser powers than suspended M-SWNTs, which can be explained by their poorer thermal conductivity than M-SWNTs.

We use five different lasers ($E_L$ = 1.579, 1.959, 2.330, 2.412, and 2.540 eV) to verify this ERS feature [Fig. 3(a)]. If our understanding to the ERS process via the high-energy $e$-$h$ pairs is correct, we expect to observe this ERS feature only with $E_L$ = 2.330, 2.412, and 2.540 eV, which satisfy the condition of $E_L \geq S_{11} + S_{22}$, but not with $E_L$ = 1.579 and 1.959 eV, which do not satisfy the condition of $E_L \geq S_{11} + S_{22}$. Experimentally, we observe various phonon Raman features of this nanotube with $E_L$ = 1.579 eV. However, we do not observe any ERS feature at $E_L - S_{22} = 0.066$ eV, which corresponds to ~530 cm$^{-1}$ (marked by a red arrow in the top panel of Fig. 3(a)). Similarly, no ERS feature is observed at $E_L - S_{22} = 0.446$ eV or ~3600 cm$^{-1}$ with $E_L$ = 1.959 eV. We do observe weak but definite ERS features with the three higher energy lasers. With $E_L$ = 2.412 eV, the ERS feature is expected at $E_L - S_{22} = 0.899$ eV or ~7250 cm$^{-1}$, and is observed at 7320 cm$^{-1}$ (FWHM = 210 cm$^{-1}$). With $E_L$ = 2.540 eV, the ERS feature is expected at $E_L - S_{22} = 1.027$ eV or ~8280 cm$^{-1}$, and is observed at 8317 cm$^{-1}$ (FWHM = 310 cm$^{-1}$). Converting the Raman shifts to the corresponding scattered photon energy [Fig. 3(b)], all the ERS features remain centered around ~1.51 eV regardless of $E_L$, which provides direct information about the $S_{22}$ of (15,2). The slightly shifted $S_{22}$ values (less than 10 meV) are caused by the temperature-induced thermal effect [34,35] and the consequent desorption of adsorbed gas molecules [36,37] on the SWNT walls at different experimental conditions including temperature variation and laser heating effect. No phonon peak is observed near the ERS features with $E_L$ = 2.412 or 2.540 eV due to the absence of phonon modes in scattered resonance condition with the $S_{22}$ at these two excitations. In particular, the fact that this feature does not appear at $S_{22}$ but a PL peak appears at $S_{11}$ when the condition of $S_{22} < E_L < S_{11} + S_{22}$ is satisfied, i.e., as shown in the spectra with $E_L$ = 1.580 eV, rules out high energy fluorescence as a possible interpretation and demonstrates the importance of high-energy e-h pairs in this process.

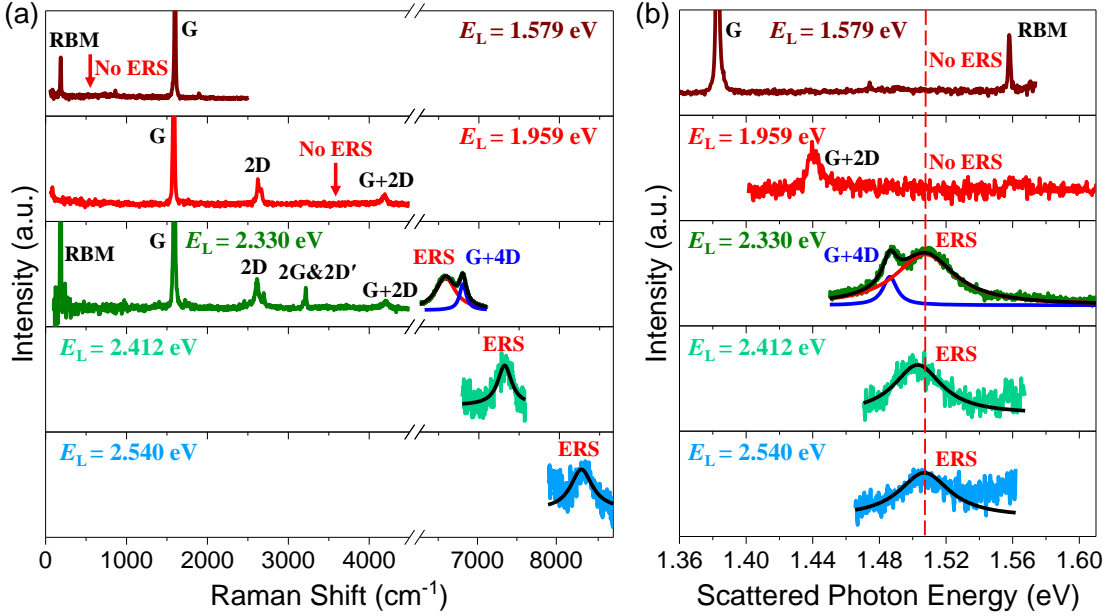

FIG. 3 Raman spectra and the ERS feature of a suspended (15,2) nanotube at different excitation lasers ($E_L$ = 1.579, 1.959, 2.330, 2.412, and 2.540 eV) with respect to (a) the Raman shift and (b) scattered photon energy. No ERS is observed with $E_L$ = 1.579 and 1.959 eV as the condition of $E_L \geqslant S_{11} + S_{22}$ = 2.325 eV is not satisfied. The ERS feature is observed at $S_{22}$ with $E_L$ = 2.330, 2.412, and 2.540 eV since the condition of $E_L \geqslant S_{11} + S_{22}$ = 2.325 eV is satisfied.

We also observe the ERS features of the other chiralities in the $2n+m$ = 32 family, to which (15,2) belongs. Figure 4 shows the ERS and PL spectra of suspended (14,4), (13,6), and (12,8) nanotubes and Table 1 summarizes all the information. In the $2n+m$ = 32 family, $S_{22}$ decreases with increasing $\theta$ [10,38]. The condition of $E_L \geq S_{11} + S_{22}$ is satisfied with $E_L$ = 2.330 eV but not with $E_L$ = 1.579 eV for all the chiralities in this family. Therefore, we expect to observe continuously redshifted ERS features (or continuously upshifted Raman shifts for the ERS features) with increasing $\theta$ at $E_L$ = 2.330 eV but no ERS feature at $E_L$ = 1.579 eV. Experimentally, we do not observe any ERS feature with the 1.579 eV laser. We observe the ERS features at 1.475, 1.415, and 1.356 eV for (14,4), (13,6), and (12,8), respectively, with the 2.330 eV laser. The (14,4) and (13,6) samples are bundled nanotubes showing additional RBM of another nanotube. This dielectric screening effect caused by nanotube bundles [18,39] also explains the redshifted transition energies of these two samples, for example, the PL peak of (14,4) is redshifted by 24 meV from the $S_{11}$ value calculated with the (*n,m*) calculator. In addition, although no ERS feature is observed at $E_L$ = 1.579 eV, the relative phonon peak intensity variations match the corresponding scattered resonance with $S_{22}$. For instance, (13,6) shows strong intermediate frequency mode (IFM, 1000-1200 cm$^{-1}$) because the scattered photon energy of IFM is close to its $S_{22}$ (~1320 cm$^{-1}$) whereas intense M-band (~1750 cm$^{-1}$) and iTOLA (~1950 cm$^{-1}$) modes are present for (12,8) with an upshifted $S_{22}$ value (~1800 cm$^{-1}$) than (13,6).

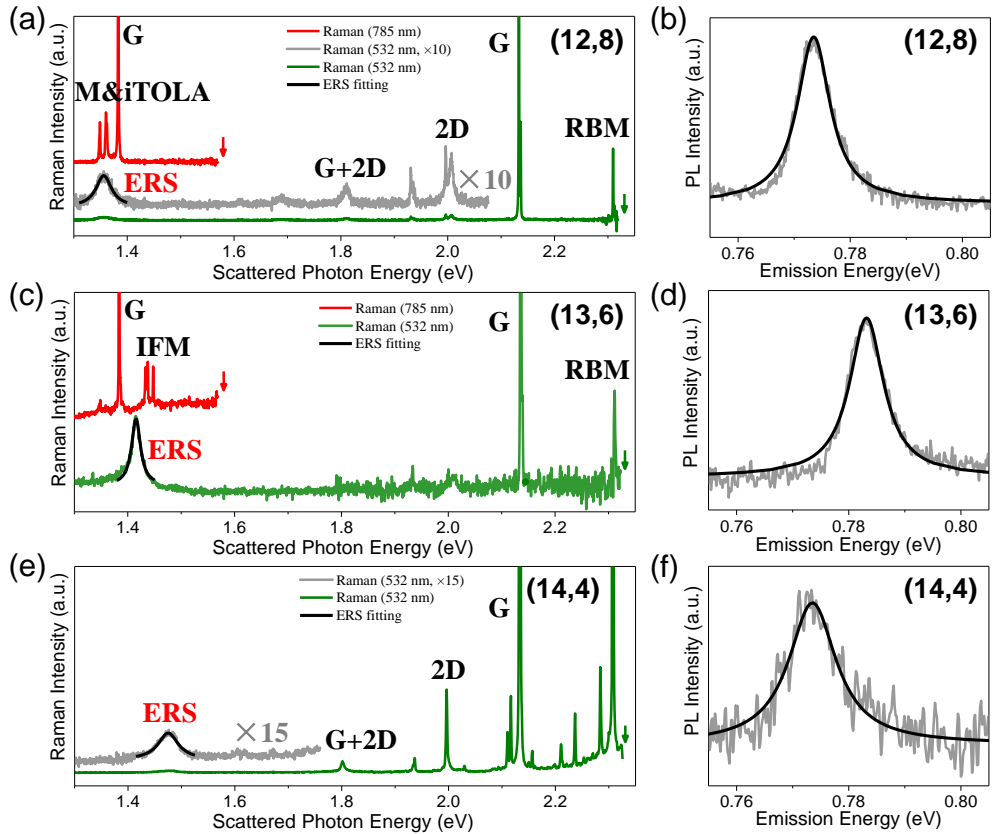

FIG. 4 Raman and ERS features (a, c, e) and PL spectra (b, d, f) of suspended (12,8), (13,6), and (14,4) tubes in the 2*n*+*m* = 32 family at 532 (green) and 785 (red) nm excitations. The ERS bands and the PL peaks are fitted with a Lorentzian line shape. The arrows indicate the energies of the corresponding excitation laser lines.

Table 1. The transition energies, ERS features, and PL peak of the suspended chiral nanotubes in the $2n+m = 32$ family.

| $(n,m)$ | $S_{11}$ [a] (eV) | $S_{22}$ [a] (eV) | $S_{11}+S_{22}$ (eV) | $E_L$ (eV) | ERS expected (eV) | ERS observed (eV) | $\Delta S_{22}$ (eV) | PL observed (eV) | $\Delta S_{11}$ (eV) |
|---|---|---|---|---|---|---|---|---|---|
| (15,2) | 0.803 | 1.522 | 2.325 | 1.579 | No ERS | None | --- | 0.799 | -0.004 |
|  |  |  |  | 1.959 | No ERS | None | --- |  |  |
|  |  |  |  | 2.330 | 1.522 | 1.513 | -0.009 |  |  |
|  |  |  |  | 2.412 | 1.522 | 1.504 | -0.018 |  |  |
|  |  |  |  | 2.540 | 1.522 | 1.509 | -0.013 |  |  |
| (14,4)[b] | 0.797 | 1.485 | 2.282 | 1.579 | No ERS | --- | --- | 0.773 | -0.024 |
|  |  |  |  | 2.330 | 1.485 | 1.475 | -0.010 |  |  |
| (13,6)[b] | 0.785 | 1.431 | 2.216 | 1.579 | No ERS | None | --- | 0.783 | -0.002 |
|  |  |  |  | 2.330 | 1.431 | 1.415 | -0.016 |  |  |
| (12,8) | 0.769 | 1.365 | 2.134 | 1.579 | No ERS | None | --- | 0.774 | 0.005 |
|  |  |  |  | 2.330 | 1.365 | 1.356 | -0.009 |  |  |

[a] Calculated using the $(n,m)$ calculator on www.chem.pku.edu.cn/cnt_assign.
[b] Bundled samples with lower transition energies than individual nanotubes.

If the condition of $E_L \geq 2 S_{11}$ is satisfied, we expect to observe an ERS feature at $S_{11}$ by the continuum of e-h pairs near $E_{e-h} = E_L - S_{11} \geq S_{11}$. However, this feature could not be distinguished from the PL peak at ambient conditions, and is not the focus of this work. We speculate that two possible methods can be used to resolve this ERS feature from the PL peak. One is by time-resolved measurements since the exciton relaxation that PL involves is a slower process than the ERS process [40]. The other is by low temperature experiments because PL corresponds to the transition from the bright excitonic states with slightly higher energy than the dark excitonic states that can also be accessed by the ERS process [41].

In conclusion, we have shown that the ERS features can be observed for S-SWNTs, which are accomplished via the available high-energy e-h pairs with $E_{e-h} \geq S_{11}$. When the condition of $E_L \geq S_{11} + S_{22}$ is satisfied, the ERS band is observed at the corresponding $S_{22}$ via those high-energy e-h pairs ($E_{e-h} = E_L - S_{22} \geq S_{11}$) with a high accuracy of ±1 meV. The ERS approach can facilitate the precise assignment of SWNT chirality and further systematic studies on the properties of SWNT with defined chirality.

The authors acknowledge the National Key Research and Development Program of China (Grants 2016YFA0201904 and 2016YFA0301204), National Natural Science Foundation of China (Grants 21873008, 21631002, 21621061 and 11874350), and Beijing National Laboratory for Molecular Sciences (BNLMS-CXTD-202001).

The authors declare no competing interests.